\documentclass{article}
\usepackage{spconf,graphicx}

\usepackage{enumitem}
\usepackage{amsmath}
\usepackage{xcolor}
\usepackage{hyperref}
\hypersetup{
    colorlinks=true,
    linkcolor=black,
    urlcolor=black,
    citecolor=black
}
\usepackage{booktabs}
\usepackage{multirow}
\usepackage{caption}
\usepackage{subcaption}
\usepackage{makecell}
\usepackage{textpos}
\usepackage{cleveref}
\usepackage[numbers, sort&compress]{natbib}

\usepackage{amssymb}
\usepackage{pifont}
\title{Multi-Scale Sub-Band Constant-Q Transform Discriminator \\ for High-Fidelity Vocoder}
\name{Yicheng Gu \qquad Xueyao Zhang \qquad Liumeng Xue \qquad Zhizheng Wu}
\address{School of Data Science, The Chinese University of Hong Kong, Shenzhen (CUHK-Shenzhen), China}
\begin{document}
\ninept
%
\maketitle
\begin{abstract}


Generative Adversarial Network (GAN) based vocoders are superior in inference speed and synthesis quality when reconstructing an audible waveform from an acoustic representation. This study focuses on improving the discriminator to promote GAN-based vocoders. Most existing time-frequency-representation-based discriminators are rooted in Short-Time Fourier Transform (STFT), whose time-frequency resolution in a spectrogram is fixed, making it incompatible with signals like singing voices that require flexible attention for different frequency bands. Motivated by that, our study utilizes the Constant-Q Transform (CQT), which owns dynamic resolution among frequencies, contributing to a better modeling ability in pitch accuracy and harmonic tracking. Specifically, we propose a Multi-Scale Sub-Band CQT (MS-SB-CQT) Discriminator, which operates on the CQT spectrogram at multiple scales and performs sub-band processing according to different octaves. Experiments conducted on both speech and singing voices confirm the effectiveness of our proposed method. Moreover, we also verified that the CQT-based and the STFT-based discriminators could be complementary under joint training. Specifically, enhanced by the proposed MS-SB-CQT and the existing MS-STFT Discriminators, the MOS of HiFi-GAN can be boosted from 3.27 to 3.87 for seen singers and from 3.40 to 3.78 for unseen singers.

\end{abstract}
\section{Introduction}
\label{sec:intro}

\begin{figure*}[htp]
    \centering
    \includegraphics[width=\textwidth]{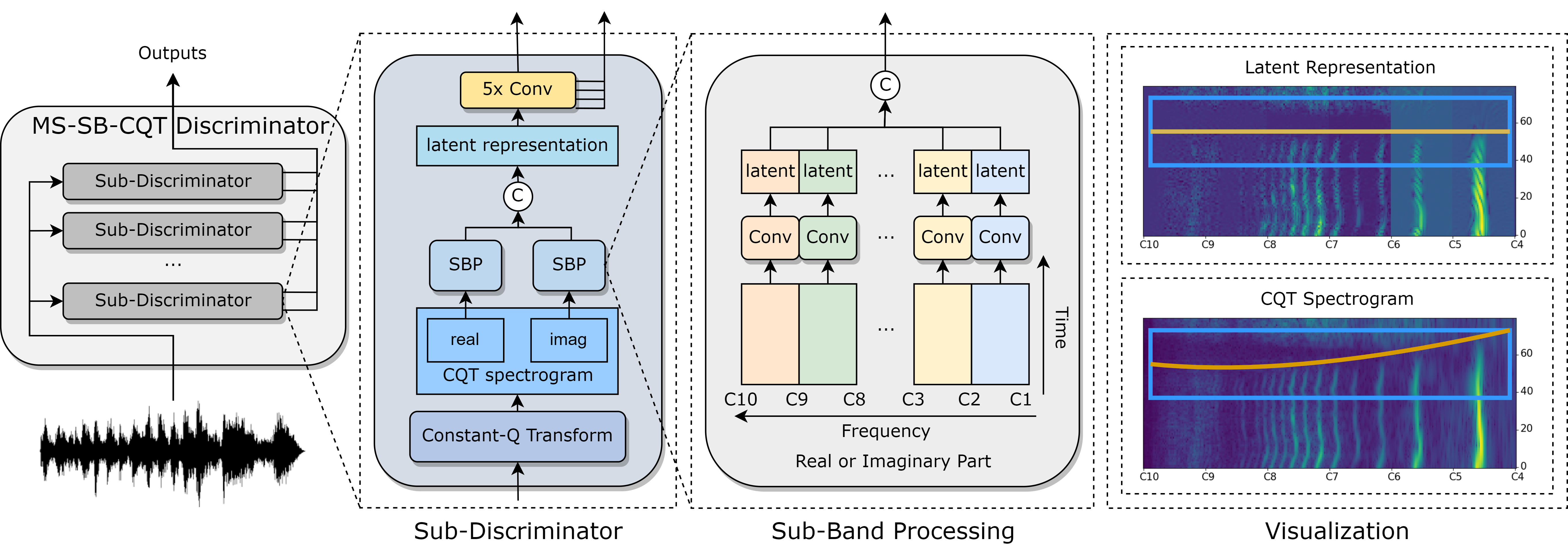}
    \caption{Architecture of the proposed Multi-Scale Sub-Band Constant-Q Transform (MS-SB-CQT) Discriminator, which can be integrated with any GAN-based vocoder. Operator ``C'' denotes for concatenation. SBP means our proposed Sub-Band Processing module. It can be observed that the desynchronized CQT Spectrogram (bottom-right) has been synchronized (upper-right) after SBP.}
    \label{fig:model}
\end{figure*}


A neural vocoder reconstructs an audible waveform from an acoustic representation. Deep generative models including autoregressive-based~\cite{WaveNet, WaveRNN}, flow-based~\cite{WaveGlow, WaveFlow}, GAN-based~\cite{PWG, MelGAN, UniversalMelGAN, HiFiGAN, FreGAN, SingGAN, BigVGAN}, and diffusion-based ~\cite{WaveGrad, DiffWave} models have been successful for this task. Because of the superior inference speed and synthesis quality, GAN-based vocoders are always attractive to researchers. However, to synthesize expressive speech or singing voice, current GAN-based vocoders still hold problems like spectral artifacts such as hissing noise~\cite{FreGAN} and loss of details in mid and low-frequency parts~\cite{SingGAN}.

To pursue high-quality GAN-based vocoders, the existing studies aim to improve both the generator and the discriminator. For the generator, SingGAN~\cite{SingGAN} adopts a neural source filter~\cite{NSF} module to utilize the sine excitation. BigVGAN~\cite{BigVGAN} introduces a new activation function with anti-aliasing modules. For the discriminator, MelGAN~\cite{MelGAN} employs a time-domain-based discriminator that successfully models waveform structures at different scales for the first time. HiFi-GAN~\cite{HiFiGAN} extends it with a Multi-Scale Discriminator and Multi-Period Discriminator, and Fre-GAN~\cite{FreGAN} further improves it by replacing the averaging pooling with discrete-wavelet-transform-based filters to preserve frequency information. UniversalMelGAN~\cite{UniversalMelGAN} introduces a Multi-Resolution Discriminator, followed by~\cite{MRD} emphasizing its significance. Encodec~\cite{encodec} extends it to the Multi-Scale STFT (MS-STFT) Discriminator.

This study focuses on improving the discriminator. Among the existing works, most time-frequency-representation-based discriminators are rooted in Short-Time Fourier Transform (STFT)~\cite{UniversalMelGAN, encodec, MRD}, which could fast extract easy-to-handle STFT spectrograms for neural networks. However, it also has limitations. Specifically, an STFT spectrogram has a fixed time-frequency resolution across all frequency bins (Section~\ref{sec:methodology-constant-q-transform}). When encountering signals like singing voices, which require different attention for different frequency bands~\cite {multisinger}, only an STFT spectrogram will be insufficient.

Motivated by that, this paper proposes a Constant-Q Transform (CQT)~\cite{CQT1992} based discriminator. The reason is that CQT has a more flexible resolution for different frequency bands than STFT. In the low-frequency band, CQT has a higher \textit{frequency resolution}, which can model the pitch information accurately. In the high-frequency band, CQT has a higher \textit{time resolution}, which can track the fast-changing harmonic variations. In addition, the CQT has log-scale distributed center frequencies, which can bring better pitch-level information~\cite{mert}. Specifically, we design a Multi-Scale Sub-Band CQT (MS-SB-CQT) Discriminator. The discriminator operates on CQT spectrograms at different scales and performs sub-band processing according to the octave information of the CQT spectrogram. Moreover, during the experiments, we find that the proposed MS-SB-CQT and MS-STFT~\cite{encodec} Discriminators can be jointly used to boost the generator further, which reveals the complementary role between the CQT-based and the STFT-based discriminators.

\section{Multi-Scale Sub-Band Constant-Q Transform Discriminator}
\label{sec:methodology}


The architecture of the proposed MS-SB-CQT Discriminator, which can be integrated into any GAN-based vocoders, is illustrated in Fig.~\ref{fig:model}. It consists of identically structured sub-discriminators operating on CQT spectrograms in different scales. Each sub-discriminator will first send the real and imaginary parts of CQT to our proposed Sub-Band Processing (SBP) module individually to get their latent representations. These two representations will then be concatenated and sent to convolutional layers to get the outputs for computing loss. The details of each module will be introduced as follows. 

\subsection{The Strengths of Constant-Q Transform}
\label{sec:methodology-constant-q-transform}
In this section, the strengths of CQT will be exhibited by introducing its design idea. We will see how CQT owns a flexible time-frequency resolution and why it can model pitch-level information better.

Following~\cite{CQT1992}, the CQT $\boldsymbol{X} ^ {cq} (k, n) $ can be defined as:
\begin{equation}
\begin{split}
\label{equa:cqt}
    \boldsymbol{X} ^ {cq} (k, n) = \sum_{j = n - \lfloor N _ {k} / 2 \rfloor }^{n + \lfloor N _ {k} / 2 \rfloor} x (j) a _ {k} ^ \ast (j - n + N _ {k} / 2),
\end{split}    
\end{equation}
where $k$ is the index of frequency bin, $x(j)$ is the $j$-th sample point of the analyzed signal, $N_{k}$ is the window length, $a_{k} (n)$ is a complex-valued kernel, and $a _ {k} ^ \ast (n)$ is the complex conjugate of $a _ {k} (n)$. 

The kernels $a _ {k} (n)$ can be obtained as:
\begin{equation}
\begin{split}
\label{eq:frame}
    a _ {k} (n) = \frac{1}{N _ {k}} w \left(\frac{n}{N _ {k}}\right) e ^ {-i2 \pi n \frac{Q _ {k}}{N _ {k}}},
\end{split}
\end{equation}
where $w(t)$ is the window function, and $Q _ {k}$ is the constant Q-factor:
\begin{equation}
\label{eq:Qfactor}
    Q _ {k} \overset{\textit{ref.}}{=} \frac{f _ {k}}{\Delta f _ {k}} = (2 ^ {\frac{1}{B}} - 1) ^ {-1},
\end{equation}
where $f _ {k}$ is the center frequency, $\Delta f _ {k}$ is the bandwidth determining the resolution trade-off, and $B$ is the number of bins per octave. 

Notably, for STFT, the $\Delta f _ {k}$ is constant, meaning the time-frequency resolution is fixed for all frequencies. However, for CQT, its main idea is to keep $Q _ {k}$ constant. As a result, the low-frequency bands will have a smaller $\Delta f _ {k}$, bringing a higher \textit{frequency resolution}, which could model the pitch information better. Besides, the high-frequency bands will have a bigger $\Delta f _ {k}$, bringing a higher \textit{time resolution}, which could track fast-changing harmonics variations better.

In Eq.~(\ref{eq:frame}), the window length of the $k$-th frequency bin, $N_k$, can be obtained as:
\begin{equation}
\label{eq:bandwidth}
    N_k = \frac{f_s}{\Delta f _ {k}} = \frac{f_s}{f _ {k}} \cdot (2 ^ {\frac{1}{B}} - 1) ^ {-1}
\end{equation}
where $f _ {s}$ is the sampling rate, and $f_{k}$ is defined as:
\begin{equation}
\label{eq:center_freq}
\begin{split}
    f _ {k} = f _ {1} \cdot 2 ^ {\frac{k - 1}{B}},
\end{split}
\end{equation}
where $f _ {1}$ is the lowest center frequency, which is set to 32.7 Hz (C1) in our study. 

\begin{table*}[t]
\centering
\caption{Analysis-synthesis results of different discriminators when being integrated into  HiFi-GAN~\cite{HiFiGAN}. The best and the second best results of every column (except those from Ground Truth) in each domain (speech and singing voice) are \textbf{bold} and \textit{italic}. ``S" and ``C" represent MS-STFT and MS-SB-CQT Discriminators respectively. The MOS scores are with 95\% Confidence Interval (CI).}\label{tab:results-effectiveness}. 
\resizebox{0.9\textwidth}{!}{%
\small
\begin{tabular}{clcccccccccc}
\toprule
   \multirow{2}{*}{\textbf{Domain}} & \multirow{2}{*}{\textbf{System}} & \multicolumn{2}{c}{\textbf{MCD ($\downarrow$)}} & \multicolumn{2}{c}{\textbf{PESQ ($\uparrow$)}} & \multicolumn{2}{c}{\textbf{FPC ($\uparrow$)}} & \multicolumn{2}{c}{\textbf{F0RMSE ($\downarrow$)}} & \multicolumn{2}{c}{\textbf{MOS ($\uparrow$)}}\\ \cmidrule(lr){3-4} \cmidrule(lr){5-6} \cmidrule(lr){7-8} \cmidrule(lr){9-10} \cmidrule(lr){11-12}
  & & \textbf{Seen} & \textbf{Unseen} & \textbf{Seen} & \textbf{Unseen} & \textbf{Seen} & \textbf{Unseen} & \textbf{Seen} & \textbf{Unseen} & \textbf{Seen} & \textbf{Unseen} \\
\midrule
\multirow{5}{*}{\makecell{\textbf{Singing}\\ \textbf{voice}}}
& Ground Truth & 0.00 & 0.00 & 4.50 & 4.50 & 1.000 & 1.000 & 0.00 & 0.00 & 4.85 $\pm$ 0.06 & 4.73 $\pm$ 0.09 \\
\cmidrule(lr){2-12}
 & HiFi-GAN & \textit{2.82} & \textit{3.17} & 2.94 & 2.86 & 0.954 & 0.961 & 56.96 & 59.28 & 3.27 $\pm$ 0.16 & 3.40 $\pm$ 0.15 \\
 & HiFi-GAN (+S) & 2.97 & 3.37 & 2.95 & 2.87 & 0.967 & 0.968 & 39.06 & 46.49 & 3.42 $\pm$ 0.16 & 3.56 $\pm$ 0.17 \\
 & HiFi-GAN (+C) & 2.90 & 3.35 & \textit{3.03} & \textit{2.95} & \textit{0.970} & \textit{0.971} & \textit{35.57} & \textit{41.09} & \textit{3.66} $\pm$ \textit{0.14} & \textit{3.63} $\pm$ \textit{0.16} \\
 & HiFi-GAN (+S+C)& \textbf{2.54} & \textbf{3.08} & \textbf{3.09} & \textbf{2.98} & \textbf{0.971} & \textbf{0.973} & \textbf{35.45} & \textbf{39.90} & \textbf{3.87} $\pm$ \textbf{0.14} & \textbf{3.78} $\pm$ \textbf{0.12} \\
\midrule
\multirow{5}{*}{\textbf{Speech}}
& Ground Truth & 0.00 & 0.00 & 4.50 & 4.50 & 1.000 & 1.000 & 0.00 & 0.00 & 4.62 $\pm$ 0.11 & 4.59 $\pm$ 0.11 \\ 
\cmidrule(lr){2-12}
 & HiFi-GAN & \textit{3.21} & 2.10 & 3.01 & 3.14 & \textit{0.883} & \textit{0.781} & 186.19 & \textit{293.34} & 3.91 $\pm$ 0.17 & 3.96 $\pm$ 0.16 \\
 & HiFi-GAN (+S) & 3.47 & 2.10 & 2.97 & 3.09 & 0.869 & 0.772 & 195.05 & 298.53 & \textbf{4.02} $\pm$ \textbf{0.15} & 4.00 $\pm$ 0.17 \\
 & HiFi-GAN (+C) & 3.26 & \textit{2.07} & \textit{3.04} & \textbf{3.16} & \textbf{0.884} & 0.768 & \textbf{180.29} & 301.83 & 4.01 $\pm$ 0.15 & \textit{4.13} $\pm$ \textit{0.14} \\
 & HiFi-GAN (+S+C)& \textbf{3.13} & \textbf{2.05} & \textbf{3.05} & \textit{3.15} & \textit{0.883} & \textbf{0.792} & \textit{182.04} & \textbf{281.90} & \textbf{4.02} $\pm$ \textbf{0.17} & \textbf{4.14} $\pm$ \textbf{0.15} \\
\bottomrule

\end{tabular}%
}
\end{table*} 

\subsection{Multi-Scale Sub-Discriminators}
\label{sec:methodology-multi-scale}
To capture the information under more diverse time-frequency resolutions, we leverage the multi-scale idea~\cite{HiFiGAN,MRD} and adopt sub-discriminators on CQTs with different overall resolution trade-offs.

Given Eq.~(\ref{eq:Qfactor}) and (\ref{eq:center_freq}), we can observe that the bandwidth $\Delta f_k$, which determines the resolution trade-off, is dependent on the number of bins per octave $B$. In other words, we can set the different $B$ to obtain the different resolution distributions. Based on that, we follow~\cite{MRD,encodec} to apply three sub-discriminators with $B$ equals 24, 36, and 48, respectively.




\subsection{Sub-Band Processing Module}
\label{sec:methodology-sub-band}



As two sides of a coin, although the dynamic bandwidth $\Delta f_k$ brings flexible time-frequency resolution, it also brings the unfixed window length $N_k$. As a result, the kernels $a_{k} (n)$ in different frequency bins are not temporally synchronized~\cite{CQT2010}. The CQT spectrogram with such artifacts has been visualized in the bottom right of Fig.~\ref{fig:model}.


To alleviate this problem, \cite{CQT2010} designs a series of kernels that are temporally synchronized within an octave. This algorithm has also been used in toolkits like librosa~\cite{librosa} and nnAudio~\cite{nnaudio}. However, such an algorithm only makes the $a_{k} (n)$ of \textit{intra-octave} temporally synchronized but leaves those of \textit{inter-octave} unsolved. During experiments, we found that just using CQT spectrograms with such a bias could even hurt the quality of vocoders (Section~\ref{sec:ablation}).



Based on that, we utilize the philosophy of representation learning and design the Sub-Band Processing (SBP) module to address this problem further. In particular, the real or imaginary part of a CQT spectrogram will first be split into sub-bands according to octaves. Then, each band will be sent to its corresponding convolutional layer to get its representation. Finally, we concatenate the representations from all bands to obtain the latent representation of the CQT spectrogram. In the upper right of Fig.~\ref{fig:model}, it can be observed that our proposed SBP successfully learns the temporally synchronized representations among all the frequency bins.



\subsection{Integration with GAN-based Vocoder}

Our proposed discriminator can be easily integrated with existing GAN-based vocoders without interfering with the inference stage. We take HiFi-GAN~\cite{HiFiGAN} as an example. HiFi-GAN has a generator $G$ and multiple discriminators $D _ {m}$. The generation loss $\mathcal{L} _ {G}$, and discrimination loss $\mathcal{L} _ {D}$ are defined as, $ \mathcal{L} _ {G} = \sum_{m = 1} ^ {M} [\mathcal{L} _ {adv} (G; D _ {m}) + 2 \mathcal{L} _ {fm} (G; D _ {m})] + 45 \mathcal{L} _ {mel},  \mathcal{L} _ {D} = \sum_{m = 1} ^ {M} [\mathcal{L} _ {adv} (D _ {m}; G), $ where M is the number of discriminators, $D _ {m}$ denotes the m-th discriminator, $\mathcal{L} _ {adv}$ is the adversarial GAN loss, $\mathcal{L} _ {fm}$ is the feature matching loss, and $\mathcal{L} _ {mel}$ is the mel spectrogram reconstruction loss. Among these losses, only $\mathcal{L} _ {fm}$ and $\mathcal{L} _ {adv}$ are related to our discriminator. Thus, just adding $\mathcal{L} _ {adv} (G; D _ {\text{MS-SB-CQT}}) + 2 \mathcal{L} _ {fm} (G; D _ {\text{MS-SB-CQT}})$ to $\mathcal{L} _ {G}$ and $\mathcal{L} _ {adv} (D _ {\text{MS-SB-CQT}}; G)$ to $\mathcal{L} _ {D}$ can integrate the proposed discriminator in the training process. 

\section{Experiments}
\label{sec:experiments}

\begin{table*}[ht]
\centering
\caption{Analysis-synthesis results of our proposed MS-SB-CQT Discriminator when integrating in MelGAN~\cite{MelGAN} and NSF-HiFiGAN in singing voice datasets. The improvements are shown in \textbf{bold}. ``S" and ``C" represent MS-STFT and MS-SB-CQT Discriminators respectively. All the improvements in MCD, PESQ, and Preference are significant ($p$-value $<$ 0.01).}\label{tab:results-sota}
\resizebox{0.8\textwidth}{!}{%
\small
\begin{tabular}{lcccccccccccccccc}
\toprule
 \multirow{2}{*}{\textbf{System}} & \multicolumn{2}{c}{\textbf{MCD ($\downarrow$)}} & \multicolumn{2}{c}{\textbf{PESQ ($\uparrow$)}}  & \multicolumn{2}{c}{\textbf{FPC ($\uparrow$)}} & \multicolumn{2}{c}{\textbf{F0RMSE ($\downarrow$)}} & \multicolumn{2}{c}{\textbf{Preference ($\uparrow$)}}\\ \cmidrule(lr){2-3} \cmidrule(lr){4-5} \cmidrule(lr){6-7} \cmidrule(lr){8-9} \cmidrule(lr){10-11}
  & \textbf{Seen} & \textbf{Unseen} & \textbf{Seen} & \textbf{Unseen} & \textbf{Seen} & \textbf{Unseen} & \textbf{Seen} & \textbf{Unseen} & \textbf{Seen} & \textbf{Unseen} \\
\midrule
Ground Truth & 0.00 & 0.00 & 4.50 & 4.50 & 1.000 & 1.000 & 0.00 & 0.00 & / & / \\ 
\midrule
 MelGAN & 4.44 & 5.21 & 2.23 & 2.15 & 0.968 & 0.964 & 46.80 & 51.73 & 8.47\% & 27.45\% \\
 MelGAN (+S+C) & \textbf{4.08}  & \textbf{4.87}  & \textbf{2.35}  & \textbf{2.23} & 0.960 & 0.962 & 51.78 & \textbf{50.99} & \textbf{91.53}\%  & \textbf{72.55}\%  \\
\midrule
 NSF-HiFiGAN & 1.73 & 2.04 & 3.95 & 3.88 & 0.985 & 0.980 & 25.62 & 31.17 & 41.67\% & 29.41\% \\
 NSF-HiFiGAN (+S+C) & \textbf{1.48}  & \textbf{1.72}  & \textbf{3.98}  & \textbf{3.91} & 0.979 & \textbf{0.983} & \textbf{24.01} & 31.19 & \textbf{58.33}\% & \textbf{70.59}\%  \\
\bottomrule

\end{tabular}%
}
\end{table*}


We conduct experiments to investigate the following four questions. \textbf{EQ1}: How effective is the proposed MS-SB-CQT Discriminator? \textbf{EQ2}: Could using MS-SB-CQT and MS-STFT Discriminators jointly improve the vocoder further? \textbf{EQ3}: How generalized is the MS-SB-CQT Discriminator under different GAN-based vocoders? \textbf{EQ4}: Is it necessary to adopt the proposed SBP module? Audio samples are available on our demo site\footnote{\url{https://vocodexelysium.github.io/MS-SB-CQTD/}}.







\subsection{Experimental setup}
\label{sec:train-data}



\textbf{Dataset} \quad 
The experimental datasets contain both speech and singing voices. For the singing voice, we adopt M4Singer~\cite{M4Singer}, PJS~\cite{PJS}, and one internal dataset. We randomly sample 352 utterances from the three datasets to evaluate \textbf{\textit{seen}} singers and leave the remaining for training (39 hours). 445 samples from Opencpop~\cite{Opencpop}, PopCS~\cite{POPCS}, OpenSinger~\cite{OpenSinger}, and CSD~\cite{csd} are chosen to evaluate \textbf{\textit{unseen}} singers. For the speech, we use the train-clean-100 from LibriTTS~\cite{LibriTTS} and LJSpeech~\cite{ljspeech}. We randomly sample 2316 utterances from the two datasets to evaluate \textbf{\textit{seen}} speakers and leave the remaining for training (about 75 hours). 3054 samples from VCTK~\cite{vctk} are chosen to evaluate \textbf{\textit{unseen}} speakers.



\textbf{Implementation Details} \quad
The CNN in SBP uses a Conv2D with kernel size of (3, 9). The CNNs in each Sub-Discriminator consist of a Conv2D with kernel size (3, 8) and 32 channels, three Conv2Ds with dilation rates of [1, 2, 4] in the time dimension and a stride of 2 over the frequency dimension, and a Conv2D with kernel size (3, 3) and stride (1, 1). For CQT, the global hop length is empirically set to 256, and the waveform will be upsampled from $f _ {s}$ to $2 f _ {s}$ before the computation to avoid the $f _ {max}$ of the top octave hitting the Nyquist Frequency.



\textbf{Evaluation Metrics} \quad
For objective evaluation, we use Perceptual Evaluation of Speech Quality (PESQ)~\cite{PESQ} and Mel Cepstral Distortion (MCD)~\cite{mcd} to evaluate the spectrogram reconstruction. We use F0 Root Mean Square Error (F0RMSE) and F0 Pearson Correlation Coefficient (FPC) for evaluating pitch stability. The Mean Opinion Score (MOS) and Preference Test are used for subjective evaluation. We invited 20 volunteers who are experienced in the audio generation area to attend the subjective evaluation. Each setting in the bellowing MOS test has been graded 200 times, and each pair in the preference test has been graded 120 times.

\subsection{Effectiveness of MS-SB-CQT Discriminator (EQ1 \& EQ2)}
\label{sec:effectiveness}


To verify the effectiveness of the proposed discriminator, we take HiFi-GAN as an example and enhance it with different discriminators. The results of the analysis-synthesis are illustrated in Table~\ref{tab:results-effectiveness}. Regarding singing voice, we can observe that: (1) both HiFi-GAN (+C) and HiFi-GAN (+S) perform better than HiFi-GAN, showing the importance of time-frequency-representation-based discriminators~\cite{MRD}; (2) HiFi-GAN (+C) performs better than HiFi-GAN (+S) with a significant boost in MOS, showing the superiority of our proposed MS-SB-CQT Discriminator; (3) HiFi-GAN (+S+C) performs best both objectively and subjectively, which shows that different discriminators will have complementary information for each other, confirming the effectiveness of jointly training. A similar conclusion can be drawn for the unseen speaker evaluation of speech data.


\begin{figure}[!ht]
    \centering
    \begin{subfigure}[b]{0.46\linewidth}
         \centering
         \includegraphics[width=0.95\linewidth,height=0.82\linewidth]{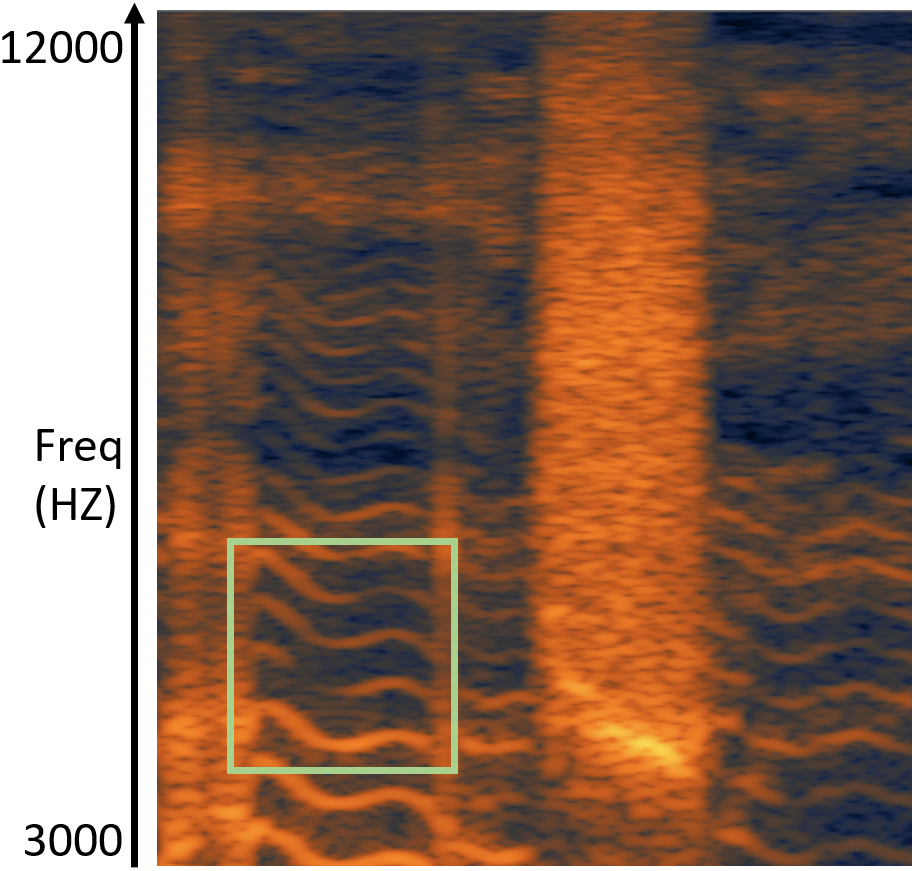}
         \caption{Ground Truth}
         \label{fig:gt}
    \end{subfigure}
    \hfill
    \begin{subfigure}[b]{0.46\linewidth}
         \centering
         \includegraphics[width=0.95\linewidth,height=0.82\linewidth]{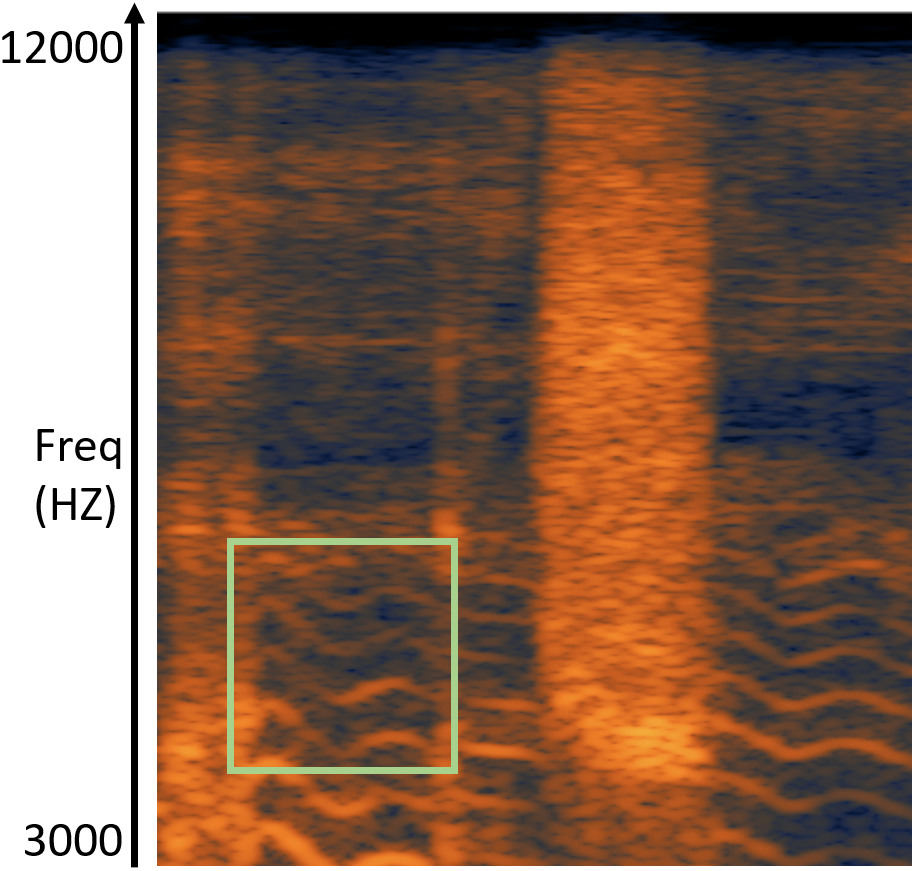}
         \caption{HiFi-GAN (+S)}
         \label{fig:stft}
    \end{subfigure} 
    \newline
    \begin{subfigure}[b]{0.46\linewidth}
         \centering
         \includegraphics[width=0.95\linewidth,height=0.82\linewidth]{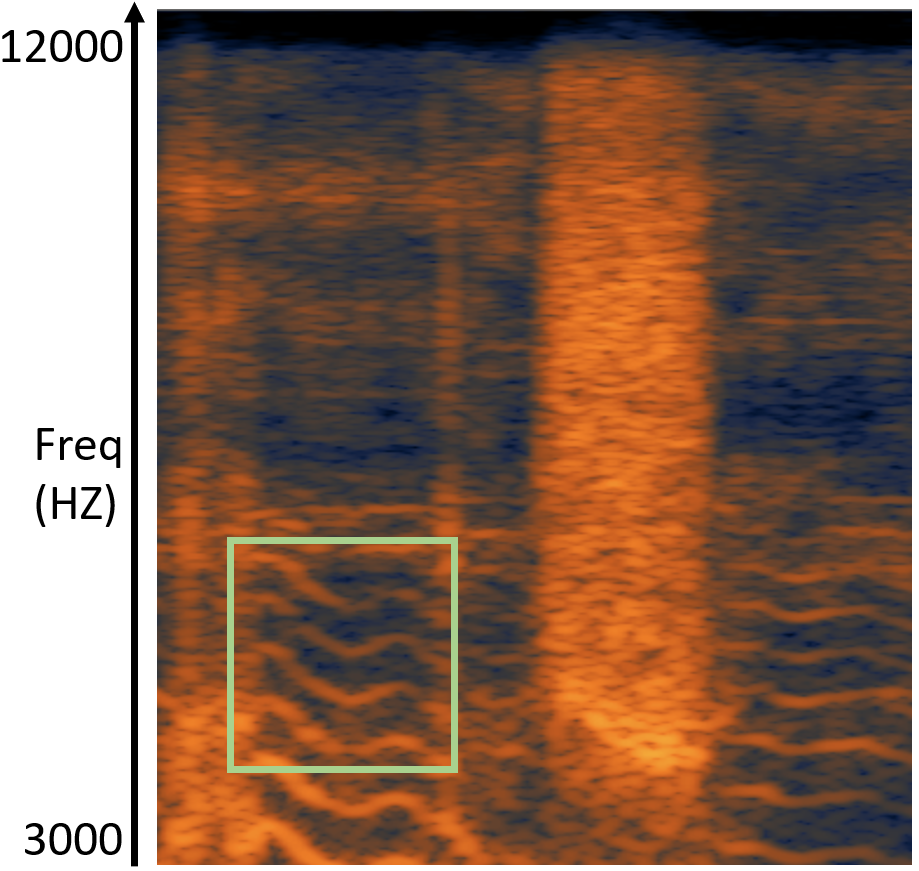}
         \caption{HiFi-GAN (+C)}
         \label{fig:cqt}
    \end{subfigure}
    \hfill
    \begin{subfigure}[b]{0.46\linewidth}
         \centering
         \includegraphics[width=0.95\linewidth,height=0.82\linewidth]{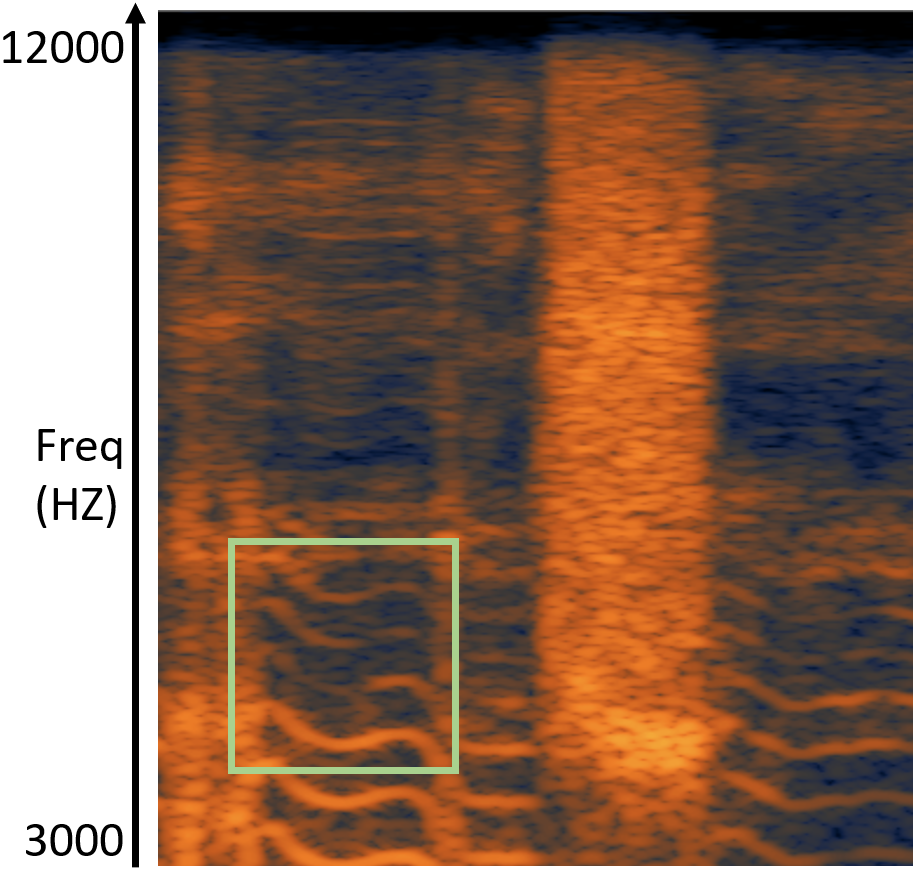}
         \caption{HiFi-GAN (+S+C)}
         \label{fig:merge}
    \end{subfigure} 
    \caption{The comparison of mel spectrograms of HiFi-GANs enhanced by different discriminators. ``S" and ``C" represent MS-STFT and MS-SB-CQT Discriminators respectively. Integrating with both CQT- and STFT-based discriminators, HiFi-GAN could achieve a higher synthesis quality with more accurate harmonic tracking and frequency reconstruction.}
    \label{fig:case-study}
\end{figure}

To further explore the specific benefits of using the CQT-based and the STFT-based discriminators jointly, we conducted a case study (Fig.~\ref{fig:case-study}). Notably, in the displayed high-frequency parts, STFT has a better frequency resolution, and CQT has a better time resolution. Regarding the frequency parts in the rectangle, it can be observed that: (1) HiFi-GAN with MS-STFT Discriminator (Fig.~\ref{fig:stft}) can reconstruct its frequency accurately but cannot track the changes due to insufficient time resolution; (2) HiFi-GAN with MS-SB-CQT Discriminator (Fig.~\ref{fig:cqt}) can track the harmonics, but the frequency reconstruction is inaccurate due to the low-frequency resolution; (3) Integrating those two combines their strengths and thus achieve a better reconstruction quality (Fig.~\ref{fig:merge}).

\subsection{Generalization Ability of MS-SB-CQT Discriminator (EQ3)}
\label{sec:training-strategy}


To verify the generalization ability of the proposed MS-SB-CQT Discriminator, besides HiFi-GAN, we also conduct experiments under MelGAN\footnote{\url{https://github.com/descriptinc/melgan-neurips/}}~\cite{MelGAN} and NSF-HiFiGAN\footnote{\url{https://github.com/nii-yamagishilab/project-NN-Pytorch-scripts}}. Note that NSF-HiFiGAN is one of the state-of-the-art vocoders for singing voice~\cite{SVCC}. It combines the neural source filter (NSF)~\cite{NSF} to enhance the generator of HiFi-GAN. The experimental results are presented in Table~\ref{tab:results-sota}. 




It is illustrated that: (1) In general, the performance of MelGAN and NSF-HiFiGAN can be improved significantly by jointly training with MS-SB-CQT and MS-STFT Discriminators, with both objective and subjective preference tests confirming the effectiveness; (2) In particular, MelGAN tends to overfit the low-frequency part and ignore mid and high-frequency components, resulting in audible metallic noise. After adding MS-STFT and MS-SB-CQT Discriminators, it could model the global information of spectrogram better\footnote{We show the representative cases on the \href{https://vocodexelysium.github.io/MS-SB-CQTD/}{demo page}.\label{demo page}}, bringing in significantly better MCD and PESQ. Although the low-frequency-related metrics worsen, the preference test shows that the overall quality has remarkably increased; NSF-HiFiGAN can synthesize high-fidelity singing voices. However, it still lacks frequency details. Adding MS-STFT and MS-SB-CQT Discriminators tackles that problem\textsuperscript{\ref{demo page}}, making synthesized samples closer to the ground truth. Subjective results with a higher preference percentage also demonstrate the effectiveness.

\subsection{Necessity of Sub-Band Processing (EQ4)}
\label{sec:ablation}

As introduced in Section~\ref{sec:methodology-sub-band}, we propose the Sub-Band Processing module to obtain the temporally synchronized CQT latent representations. To verify the necessity of it, we conduct an ablation study that removes the SBP module from the proposed MS-SB-CQT Discriminator. We adopt Opencpop~\cite{Opencpop} as the experimental dataset. We randomly selected 221 utterances for evaluation and the remaining for training (about 5 hours).

\begin{table}[ht]
\caption{Analysis-synthesis results of HiFi-GAN enhanced by different CQT-based discriminators. MS-CQT Discriminator represents a discriminator that only removes the Sub-Band Processing module from our proposed MS-SB-CQT Discriminator. }\label{tab:results-ablation}
\centering
\resizebox{\linewidth}{!}{%
\small
\begin{tabular}{lcccc}
\toprule
{\textbf{System}} & \textbf{MCD ($\downarrow$)} & \textbf{PESQ ($\uparrow$)} & \textbf{FPC ($\uparrow$)} & \textbf{F0RMSE ($\downarrow$)} \\
 \midrule
HiFi-GAN & 3.443 & 2.960 & 0.972 & 40.409 \\
\quad + MS-CQT & 3.502 & 2.932 & 0.964 & 50.918 \\
\quad + MS-SB-CQT & \textbf{3.263} & \textbf{2.985} & \textbf{0.986} &\textbf{28.313} \\
\bottomrule
\end{tabular}%
}
\end{table}


In Tabel~\ref{tab:results-ablation}, we can see that HiFi-GAN can be enhanced successfully by our proposed MS-SB-CQT Discriminator. However, just applying the raw CQT to the discriminator (MS-CQT) would even harm the quality of HiFi-GAN. We speculate this is because the temporal desynchronization in inter-octaves of the raw CQT would burden the model learning. Therefore, it is necessary to adopt the proposed SBP module for designing a CQT-based discriminator.





\section{Conclusion and Future Works}
\label{sec:conclusion-and-future-works}

This study proposed a Multi-Scale Sub-Band Constant-Q Transform (MS-SB-CQT) Discriminator for GAN-based vocoder. The proposed discriminator outperforms the existing Multi-Scale Short-Time-Fourier-Transform (MS-STFT) Discriminator on both speech and singing voice. Besides, the proposed CQT-based discriminator can complement the existing STFT-based discriminator to improve the vocoder further. In future work, we will explore more time-frequency representations and other signal-processing approaches for better discriminators or generators.




\clearpage
\section{Reference}

\renewcommand\refname{\vskip -1cm}
\bibliographystyle{IEEEbib_new}
\bibliography{refs}

\end{document}